\documentclass[10pt, conference]{IEEEtran}

\usepackage[utf8]{inputenc} 

\usepackage[american]{babel} 
\usepackage{csquotes} 
\usepackage[babel=true]{microtype} 
\usepackage{balance}

\usepackage{graphicx}
\graphicspath{{./figures/}}
\usepackage{standalone}
\addto\captionsamerican{} 
\usepackage{varioref}

\usepackage{booktabs}

\usepackage{standalone}
\usepackage{tikz}
\usetikzlibrary{calc,patterns,decorations.pathmorphing,decorations.markings}
\usetikzlibrary{positioning}
\usetikzlibrary{shapes,decorations}
\usetikzlibrary{shapes.geometric, arrows, positioning}

\usepackage{mathtools}
\usepackage{mathptmx}

\usepackage{cite}

\newcommand{\Hair}{\ifmmode\mskip1mu\else\kern0.08em\fi}

\usepackage{tikz}
\usetikzlibrary{arrows}
\usetikzlibrary{chains,shapes.arrows,fit}
\usetikzlibrary{positioning}
\usetikzlibrary{calc} 
\usetikzlibrary{circuits.logic.US,circuits.logic.IEC}
\usetikzlibrary{shapes,decorations}
\usetikzlibrary{intersections}
\usepackage{tkz-euclide}

\title{A Systems Approach for Eliciting\\ Mission-Centric Security Requirements}

\author{ \IEEEauthorblockN {
	Bryan Carter\IEEEauthorrefmark{1}\IEEEauthorrefmark{3},
	Georgios Bakirtzis\IEEEauthorrefmark{2}\IEEEauthorrefmark{4},
	Carl Elks\IEEEauthorrefmark{2}\IEEEauthorrefmark{5}, and
	Cody Fleming\IEEEauthorrefmark{1}\IEEEauthorrefmark{6}
    }
\IEEEauthorblockA {
    \IEEEauthorrefmark{1}Systems \& Information Engineering, University of Virginia, Charlottesville, VA USA\\
    \IEEEauthorrefmark{2}Electrical \& Computer Engineering, Virginia Commonwealth University, Richmond, VA USA\\
  	\IEEEauthorrefmark{3}bcarter@virginia.edu
	\IEEEauthorrefmark{4}bakirtzisg@ieee.org
	\IEEEauthorrefmark{5}crelks@vcu.edu
	\IEEEauthorrefmark{6}fleming@virginia.edu
    }
}

\date{}

\begin{document}

\maketitle

\begin{abstract}
The security of cyber-physical systems is first and foremost a safety problem, yet it is typically handled as a traditional security problem, 
which means that solutions are based on defending against threats and are often implemented too late. 
This approach neglects to take into consideration the context in which the system is intended to operate, thus system safety may be compromised. This paper presents a systems-theoretic analysis approach that combines stakeholder perspectives with a modified version of Systems-Theoretic Accident Model and Process (STAMP) that allows decision-makers to strategically enhance the safety, resilience, and security of a cyber-physical system against potential threats. This methodology allows the capture of vital mission-specific information in a model, which then allows analysts to identify and mitigate vulnerabilities in the locations most critical to mission success. 
We present an overview of the general approach followed by a real example using an unmanned aerial vehicle conducting a reconnaissance mission. 
\end{abstract}

\section{Introduction}
Assessing the security of Cyber-Physical Systems (CPS) has long been handled in the same manner as that of software security: that is to identify and address individual component threats. These threats are often identified by analysts using a variety of threat detection methodologies. However, it has become increasingly common, when dealing with complex coupled systems, that vulnerabilities are only identified during forensic analysis after a security breach~\cite{Equifax,Havex} or even after detrimental effects have already taken place~\cite{Stuxnet}. This issue is particularly concerning in the realm safety-critical CPS, where such security breaches can put human lives in immediate danger. 
This is the case due to the intrinsic interaction between software-oriented control and the physical world in CPS, where the lines between the fields of safety and security become blurred, such that it is necessary to consider them as a single entity when trying to ensure the successful and safe operation of CPS. 

An important metric that has often been neglected in the security of CPS is the specific mission, i.e., expected service, that it is intended to perform. Traditional analysis methods are mission-agnostic, vulnerabilities are viewed in the context of whether or not security is breached, regardless of the magnitude of the breach’s effect on its mission requirements or possible unacceptable outcome later on.  
By creating a mission-aware  analysis, the steps towards mitigating a vulnerability are taken in a manner that prioritizes the outcome of the mission. 
This not only includes traditional security solutions, but also resiliency solutions. Resiliency in the context of CPS refers to the ability of the system to continue to provide the expected service despite cyber attacks or other disturbances. 
This means that on the one end, a vulnerability may be ignored if it has no effect on mission outcome while on the other end, classes of vulnerabilities that could potentially disrupt the mission of the CPS might require extensive preemption and mitigation strategies.

This approach to CPS cyber-security is born out of the need to assure the successful mission of military systems such as Unmanned Aerial Vehicles (UAV), smart munitions, and other vehicles against adversaries with varying capabilities. For example, a small, hand-launched UAV used for tactical reconnaissance in Iraq likely has significantly fewer threats to mission success than that of a large, strategic reconnaissance UAV used against a nation-state. This concept extends to non-military systems as well. Autonomous vehicles have major security concerns, but the security needs may differ based on the mission it is assigned to perform. In civilian applications, an autonomous highway vehicle might have far greater potential for harming others than an autonomous farming vehicle collecting produce, thus the measures taken to secure each vehicle should differ accordingly. 
This strategy allows for informed security decisions, especially in resource-limited scenarios, and prevents securing a system from becoming 
unmanageable. 

\begin{figure}[!t]
\definecolor{acol}{RGB}{115, 115, 115}
\definecolor{c1col}{RGB}{115, 115, 115}
\definecolor{c2col}{RGB}{115, 115, 115}
\definecolor{t1col}{RGB}{237, 164, 60}
\definecolor{t1ax}{RGB}{96, 96, 96} 
\definecolor{t2col}{RGB}{20,20,155} 
\definecolor{t2fil}{RGB}{182,208,255}
\definecolor{bubcol}{RGB}{115, 115, 115}

\tikzstyle{box1}=[thick,draw=black,text=black,fill=none,minimum size=2em]
\tikzstyle{box2}=[thick,draw=red,text=black,fill=none,minimum size=2em]
\tikzstyle{vecArrow} = [very thick,draw=t2col, decoration={markings,mark=at position    1 with  {\arrow[semithick,draw=t2col,scale=2.5,fill=t2fil]{triangle 90}}},    double distance=8.4pt, shorten >= 10.5pt,    preaction = {decorate},    postaction = {draw=t2fil,line width=8.3pt,shorten >= 7.0pt}]
\tikzset{axes/.style={very thick,draw=acol,     decoration={markings,mark=at position 1 with {\arrow[scale=1.8,acol]{>}}},     postaction={decorate},     shorten >=0.4pt},   big arrow/.default=blue}
\tikzset{pointer/.style={very thick,draw=bubcol,     decoration={markings,mark=at position 1 with {\arrow[scale=1.4,bubcol]{>}}},     postaction={decorate},     shorten >=0.4pt},   big arrow/.default=blue}
\begin{tikzpicture}[node distance=2.5cm,auto,>=latex']
  \draw[axes] (-0.2,0) to (7.35,0);
  \draw[axes] (-0.2,0) to (-0.2,4.7);
  \node[rotate=90] at (-.4,2.5) {\bfseries\textcolor{t1col}{Cost}, \textcolor{t2col}{Effectiveness}};
  \draw[domain=0.5:7,smooth,variable=\x,t1col,ultra thick] plot ({\x},{0.1*exp(0.525*\x)+0.5});
  \draw[domain=0.5:7,smooth,variable=\x,t2col,ultra thick] plot ({\x},{0.1*exp(0.465*(8.5-\x))+0.5});
  \node[draw=t1col,ultra thick,fill=white,circle,minimum size=0.6cm,inner sep=0pt] at (6.6,3.75) {2};
  \node[draw=t2col,ultra thick,fill=white,circle,minimum size=0.6cm,inner sep=0pt] at (1.,3.75) {1};
  \node[align=left,text width=1.5in,anchor=west] at (1.3,4.0) {Ability to\\ impact cost \&\\ performance};
  \node[align=right,text width=1in,anchor=east] at (6.3,4.0) {Cost of\\ design\\ changes};
  \def\xlabls{{"Concept","Requirements","Design","Build","Operate"}}
  \def\xticks{{0.5,2.2,3.9,5.2,6.4}}
  \foreach \i in {0,1,2,3,4}
  {
    \pgfmathsetmacro{\x}{\xticks[\i]}
    \draw[draw=acol,very thick] (\x,-0.3) -- (\x,0.3);
    \node[text=t1ax,font=\small\bfseries] at (\x,-0.5) {\pgfmathparse{\xlabls[\i]}\pgfmathresult};
  }
  \def\hlabls{{"?","Preliminary Hazard Analysis","System \& Sub-system Hazard Analysis","Accident Analysis"}}
  \def\hticks{{1.,2.6,4.6,6.6}}
  \node[draw=bubcol,ultra thick,text width=0.75in,fill=white] (decisions) at (1.08,1.8) {80\% of\\ decisions~\cite{frola_system_1984}};
  \draw[pointer] ($(decisions.north)+(0.3,0)$) -- ($(decisions.north)+(0.3,0.75)$);
  \draw[pointer] ($(decisions.south)+(0.3,0)$) -- ($(decisions.south)+(0.3,-0.55)$);
\end{tikzpicture}
\caption{Decision effectiveness during life cycle (adapted from~\cite{strafaci_what_2008}).}
\label{fig:cost}
\end{figure}
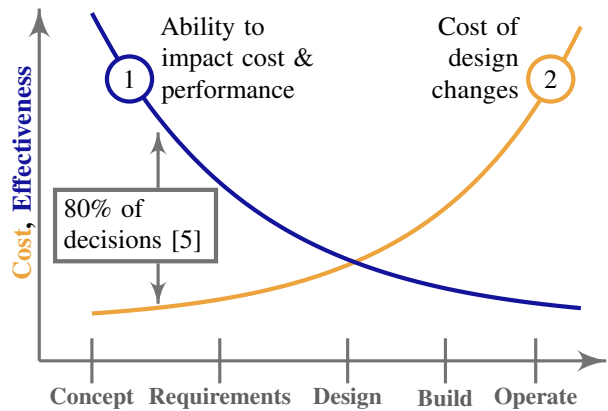

With the concepts mentioned above in mind, we propose a new, top-down analysis and modeling methodology that takes a mission-centric viewpoint to safety and security of CPS. This methodology combines inputs from system experts at the design and user levels utilizing Systems-Theoretic Accident Model and Process (STAMP)~\cite{leveson_engineering_2012} to identify potentially hazardous states that a CPS can enter and reason about how transitioning into those states can be prevented. By focusing on the intended mission, this methodology can be applied to both existing and yet-to-be designed systems, which allows for security analysis to occur earlier in the design cycle. According to Figure\vref{fig:cost}, this allows for security solutions to have both greater impact on performance and reduced cost of implementation. 
Additionally, this proactive, data-driven approach is in contrast to the reactive, ``whack-a-mole'' style of other security strategies. 

The proposed approach consists of two main parts. The first, termed the War Room, is an elicitation exercise with the main stakeholders of the mission that serves to gather mission goals, objectives, expectations, and procedures. By eliciting information from both the designers and users of the system, we attempt to minimize the disconnects between the design and actual implementation or use of the system. The second part consists of a modeling phase, which utilizes the information collected in the War Room. Specifically, the modeling phase utilizes systems and control theory to formalize the natural language outputs of the War Room and connect the abstract, mission-level information to concrete, hardware or software elements of the system. The resulting model can then be used for qualitatively identifying vulnerabilities and acting on them, or as a building block for more quantitative vulnerability and threat analysis.

The contributions of this paper are:
\begin{itemize}
\item a new analysis methodology that incorporates stakeholder perspectives to give a mission-centric viewpoint to enhancing the safety, security, and resiliency of a particular CPS;
\item a modeling technique that captures the behavior of the CPS within its mission; and 
\item a concrete application to a real-world CPS and its corresponding mission.
\end{itemize}

\section{Mission-Centric Cybersecurity}

The information carried out by both the War Room and the STPA-Sec hazard analysis not only assist us in facilitating systematic requirements and model development but, also, allow us to secure system's more effectively by being \textit{aware} of their mission-level requirements. For one, by going through the methodology above we are not blindly securing subsystems but, rather, we distill to the subsystems that are important toward a mission goal. Then, we can erect barriers in those subsystems that can assure, within error, the successful mission because we have addressed the possible insecure controls that can lead to unsafe behavior. This benefit becomes more apparent when we deal with multiple complex system's that coordinate with each other to achieve mission success. 

Traditionally, there is no ``science'' to applying security as a structured assistant to mission success. Indeed, it is often true that the procedure of securing mission-critical system's is based upon an unstructured and ultimately random security assessment that might or might not lead to mission degradation (see policy lists). This is problematic because security should not be exercised for the sake of security but, in general, should be used as a tool to avoid possible transitioning to states that violate the system's expected service. Avoiding this transitioning to hazardous states is the \textit{raison d'être} of security. Following that definition, any security measure that goes beyond providing assurance of safe behavior or any measure that doesn't adequately assure the safe behavior of the system during a mission is a loss of resources and, hence, can inadvertently be a hindrance in the command and control of military systems.

\begin{figure*}[!t]
\includegraphics[width=\textwidth]{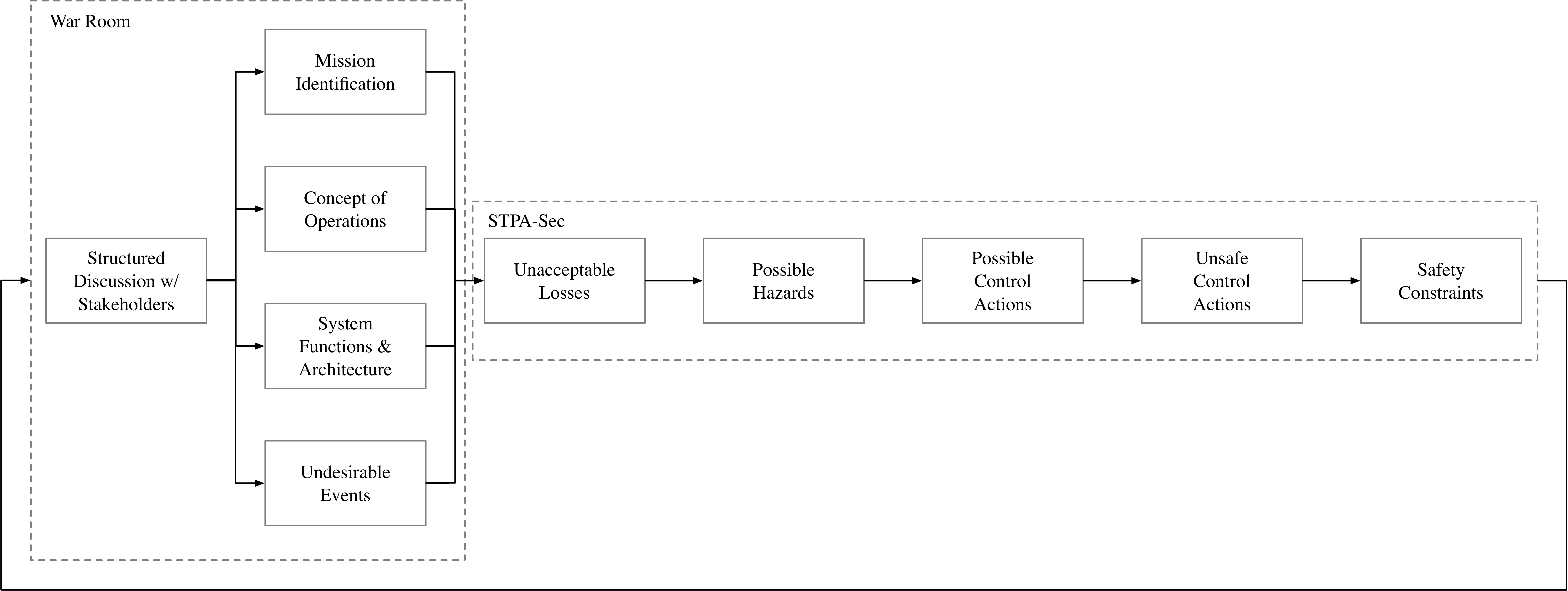}
\caption{Concept view of how the War Room facilitates the STPA-Sec analysis.}
\label{fig:concept}
\end{figure*}

\section{The War Room}
Similar to the manner in which a group of generals might plan a military operation around a large table in a meeting room, our so-called War Room aims to gather as much information as possible about a mission and the use of a CPS within that mission. This includes the objectives, success criteria, material needs, and other similar information about a mission in general, as well as the particular role a CPS would have during that mission. This exercise is completed by an analyst team leading a structured discussion with a range of stakeholders relevant to the CPS and mission. The main products are outlined in Figure\vref{fig:concept} and include a detailed, natural language description of the mission, a concept of operations (ConOps) specific to the CPS’s use in that mission, a list of functions or components that are critical to mission-success, and insights about unacceptable, hazardous, or undesirable events or outcomes with respect to the mission. This information serves as the basis for the second step of our proposed methodology; however, it is extremely valuable on its own for guiding future cybersecurity solutions.
\subsection{Conducting the War Room}

A key tenet to our approach is bringing together the system's stakeholders that can better inform the use, operation, and requirements of the mission and, consequently, the desired behavior of the system itself. For a military CPS, these stakeholders include system designers, military commanders, military operators, maintenance technicians, and potentially other personnel that might hold pertinent information to the success of the mission.

By 
discussing with the stakeholders of the mission, we are gaining an important understanding about the expectations, requirements, world-views, and interactions that each stakeholder has with the system. Since each stakeholder has a different role and area of expertise, the differing views on how the system operate and perform give more well-rounded insights and context to how a system will be used as part of a larger mission.

The analyst team is responsible for leading the discussion between the stakeholders. It is their duty to guide topics and ask specific questions that provide the information needed to construct a full model of the mission and CPS. The analysts should ensure that they have a clear understanding of the basic architecture, function, and purpose of the mission and CPS before moving on to the next phase of the discussion process. 
The analysts should obtain a list of the mission goals, sub-goals, success criteria, and reasons for mission failure. Additionally, the analysts should consult with the mission planners on what may constitute an unacceptable outcome of the mission. For example, an unacceptable outcome could include collateral damage in an air strike mission or a failure to gather information in a surveillance mission. By listing out the unacceptable outcomes of the mission, we begin to develop a sense of mission critical functions, objectives, and actions. 

After the analyst team gathers the more general information described above, the next step is to challenge the stakeholders' routines, expectations, and experiences with both the CPS and the mission. The analysts should ask questions, based on the previously gathered information, about what the stakeholder may do if a particular situation arises during the mission. For example, an analyst may ask the CPS operator what he or she might do if the CPS lost functionality during the mission. The purpose of asking these questions is two-fold: to get the stakeholders to think about how they may or may not be able to adapt to losses in functionality due to cyber events or other causes, and to further develop an understanding of the critical aspects of the mission and CPS. The answers to these questions can highlight potential oversights in the mission or the CPS, as well as further inform the construction of the model in the next steps of the methodology. 

After the War Room is completed, the logs of the discussions contain vast amounts of information that is difficult to use on its own. Consequently, it is necessary to organize that information into a more formal form, which can allow for direct analysis of the mission and CPS, in addition to providing the structure for later models used for automated vulnerability analysis.

\section{STAMP \& STPA-Sec}
To increase the interpretability of the information collected in the War Room, we propose using a modified version of STPA-Sec~\cite{young_systems_2013}, which is itself derived from STAMP~\cite{leveson_engineering_2012}.

System-Theoretic Accident Model and Processes (STAMP) is an accident causality model that captures accident causal factors including organizational structures, human error, design and requirements flaws, and hazardous interactions among non-failed components~\cite{leveson_engineering_2012}. In STAMP, system safety is reformulated as a system control problem rather than a component reliability problem—-accidents occur when component failures, external disturbances, and/or potentially unsafe interactions among system components are not handled adequately or controlled. In STAMP, the safety controls in a system are embodied in the hierarchical safety control structure, whereby commands or control actions are issued from higher levels to lower levels and feedback is provided from lower levels to higher levels. STPA-Sec is an analysis methodology based on the STAMP causality model, which is used to identify cyber vulnerabilities~\cite{young_systems_2013}.
By using this framework, we are able to capture the relevant information from the War Room in a systems-theoretic model of the mission and the CPS. This model systematically encodes the unacceptable outcomes of the mission, the hazardous states that can lead to those outcomes, and the control actions and circumstances under which those actions can create hazardous states. This information can be modeled from the mission-level all the way down to the hardware and component level, which allows for full top-to-bottom and bottom-up traceability. This traceability allows us to evaluate the cascading effects of specific changes to hardware, software, the order of operations, or other similar events on the potential outcome of a mission. Consequently, we can then use this information to identify and evaluate vulnerable areas in a system and take steps to mitigate or eliminate those vulnerabilities. 

\subsection{Constructing the Model}


The STPA-Sec model identifies how security issues can lead to accidents or unacceptable losses. In particular, the model outlines the behavior of the CPS and other actors within the overall mission and how that behavior can become unsafe. The general process for creating the STPA-Sec model is outlined in Figure\vref{fig:concept}. 


The first step of building the model is identifying the mission to be performed, which was explicitly defined in the War Room. 
This statement takes the form of, ``The mission is to perform a task, which contributes to higher-level missions or other purposes." The specific language used here serves to succinctly and precisely outline the general purpose and function of the mission. After defining the mission, the next step is to define the unacceptable outcomes or losses associated with the mission. For example, an armed UAV conducting a strike mission may have an unacceptable loss defined as any friendly casualties occurring. These losses or outcomes were either explicitly or implicitly identified and prioritized by the War Room stakeholders. For example, the failure to destroy a target may be less important to mission commanders than inflicting friendly casualties. 

After defining the unacceptable losses, we define a set of hazardous scenarios that could potentially result in an unacceptable outcome. Some of these scenarios may have been described in the War Room; however, it is likely that many will be defined by the analysts on their own. For example, in the UAV mission and unacceptable loss described above, a hazardous scenario might be that friendly forces are within the targeting area. This on its own does not necessarily lead to the unacceptable loss of friendly casualties, but such an outcome is certainly a possibility if the munition is in fact launched. The set of hazardous scenarios does not have to be an exhaustive list; however, the analysts should strive to define a set of hazards that have a reasonable chance of occurring during a mission. 

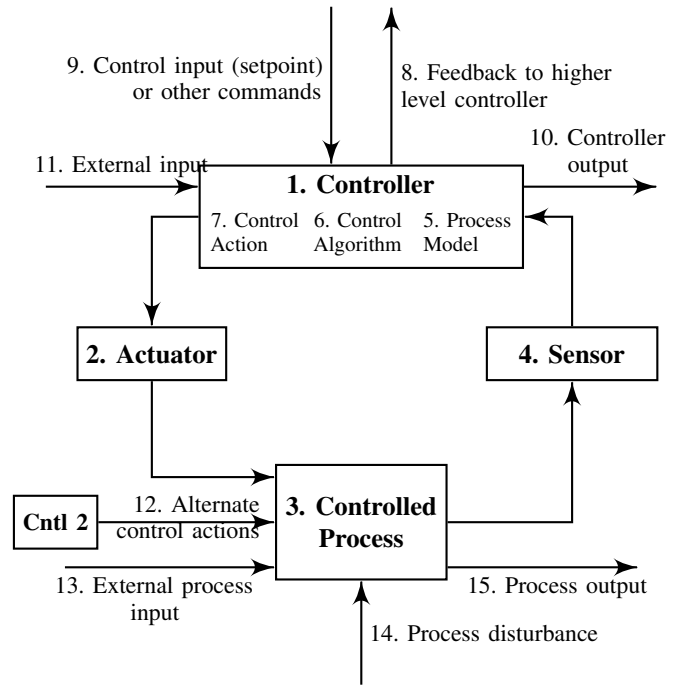
\begin{figure}[tbh]
\begin{centering}
\setlength{\tabcolsep}{1pt}
\definecolor{r1}{RGB}{0,0,0}  
\tikzstyle{box1}=[thick,draw=black,fill=none,minimum size=2em] 
\tikzstyle{box2}=[draw=black, text=red,minimum size=2em]
\tikzstyle{box3}=[draw=black, minimum size=2em]
\tikzset{connector/.style={thick,draw=r1,     decoration={markings,mark=at position 1 with {\arrow[scale=1.8,r1]{>}}},     postaction={decorate},     shorten >=0.4pt},   big arrow/.default=blue}
\begin{tikzpicture}[node distance=2.5cm,auto,>=latex']
	\node [box1] (controller)
		{
		\begin{minipage}[b]{1.6in}
			\centering{\textbf{1. Controller}}\\
			\small{}\vskip+6pt 
			\begin{tabular}{p{0.32\textwidth}p{0.34\textwidth}p{0.32\textwidth}}
			\raggedright \footnotesize{ 7. Control Action}
			& \raggedright \footnotesize{6. Control Algorithm}
			& \raggedright \footnotesize{5. Process Model}  \tabularnewline \end{tabular}
		\end{minipage}
		};
	\node [above of=controller,node distance=1.1in] (abovecntl) {};
	\node [box1,below left=0.3in and -0.15in of controller] (actuator) 
		{\textbf{2. Actuator}};
	\node [box1,below of=controller,node distance=1.6in,minimum height=0.6in] (cntlprocess) 
		{\parbox{.8in}{\centering 
		{\textbf{3. Controlled Process}}}};
	\node [below of=cntlprocess,node distance=.9in] (belowproc) {};
	\node [box1,below right=0.3in and -0.2in of controller,node distance=4.15in] (sensor) {\parbox{0.8in}{\centering {\textbf{4. Sensor}}}};
	\node [box1,left of=cntlprocess,node distance=1.6in] (controller2) 
		{\small{\textbf{Cntl 2}}};
	\node [left of=cntlprocess,node distance=1.6in] (input1) {};
	\node [right of=cntlprocess,node distance=1.4in] (output1) {};
	\node [right of=controller,node distance=1.5in] (output2) {};
	\node [left of=controller,node distance=1.7in] (input2) {};
	\draw [connector] (controller) -| (actuator);
	\draw [connector] (actuator) |-  ($(cntlprocess.west)+(0cm,0.6cm)$);
	\draw [connector] (cntlprocess) -|  (sensor);
	\draw [connector] (sensor) |- (controller);
	\draw [connector] ($(abovecntl)+(-.4cm,0)$) -- node[left,yshift=0.1cm] {\parbox{1.4in}{\raggedleft\small{9. Control input (setpoint) or other commands}}} ($(controller.north)+(-0.4cm,0)$);
	\draw [connector] ($(controller.north)+(0.4cm,0)$) -- node[right] {\parbox{1.1in}{\raggedright\small{8. Feedback to higher level controller}}} ($(abovecntl)+(.4cm,0)$);
	\draw [connector] (belowproc) -- node[right] {\parbox{1.2in}{\raggedright\small{14. Process disturbance}}} (cntlprocess);
	\draw [connector] (controller2) -- node[xshift=-0.3cm,yshift=-10pt] {\parbox{1.in}{\raggedleft\small{12. Alternate control actions}}} (cntlprocess);
	\draw [connector] ($(input1.east)-(0.,0.6cm)$) -- node[xshift=-0.2cm,below] {\parbox{1.2in}{\centering\small{13. External process input}}} ($(cntlprocess.west)-(0.cm,.6cm)$);
	\draw [connector] ($(cntlprocess.east)-(0.cm,.6cm)$) -- node[xshift=0.2cm,below] {\parbox{1.2in}{\centering\small{15. Process output}}} ($(output1.east)-(0.cm,.6cm)$);
	\draw [connector] ($(controller.east)+(0.cm,.4cm)$) -- node[xshift=0.1cm,above] {\parbox{1.in}{\centering\small{10. Controller output}}} ($(output2.east)+(0.cm,.4cm)$);
	\draw [connector] ($(input2.east)+(0.cm,.4cm)$) -- node[xshift=0.cm,above] {\parbox{.9in}{\centering\small{11. External input}}} ($(controller.west)+(0.cm,.4cm)$);
\end{tikzpicture}
\par\end{centering}
\caption{Control Loop with generic entities\label{fig:control-loop-generic-entities} (adapted from~\cite{fleming_systems_2017})}
\label{fig:loop}
\end{figure}

After defining the set of hazards, the analysts shall outline a functional hierarchy and the control actions that can be taken at each level during the mission.
For example, in a typical mission, there might be three functional levels or actors: the mission planner, the system operator, and the physical system. Obviously, the defined functional levels depend on how the analysts define them and can vary depending on the system in question, yet this step is necessary as it allows us to scope the model to a reasonable degree of granularity. 
Next, the analysts define the control actions that can be taken at each level. A generic control loop is presented in Figure\vref{fig:loop}. In general, the control action at one functional level enacts a change onto a controlled process at a lower level via an actuator and then the controller receives feedback from the controlled process via a sensor.  For example, a control action in the mission planning functional level could be defining a flight plan for an unmanned reconnaissance mission, and a control action at the operator level in the same mission might be commanding the vehicle to make a 30 degree turn to the north. The control actions and functional levels should be represented in a flow diagram that represents the planned order with respect to the mission. This will help analysts establish the ``baseline'' order of operations and procedures during the mission, which can be used later to analyze deviations from standard operating procedure.

After defining the control actions within a mission, the next step is to define the circumstances in which a particular control action could be unsafe. These circumstances can generally be defined as being a part of one of the four following categories: \begin{itemize}
\item{not providing a control action causes a hazard;}
\item{providing the control action causes a hazard;}
\item{the control action is performed at the incorrect time or out of order;} 
\item{the control action is applied for too long or stopped too soon.}
\end{itemize}
For example, in the reconnaissance mission described above, the turning control action could be unsafe when it is applied for too long and causes the aircraft to stall out. For each control action, the analysts should identify a circumstance for each of the four categories mentioned above in which the control action would be unsafe. These unsafe control actions are placed into a table, which then allows us to easily pull information when needed during the next phase of analysis. 

\subsection{Using the Model}


At the beginning of this process, several unacceptable outcomes of the mission are identified and prioritized. The next step is to identify (a) how those losses may occur and then (b) how they can be avoided or mitigated. 

In the previous step, the circumstances under which control actions would become unsafe were identified. Now, those circumstances are used to derive security requirements and constraints on the behavior of the system. For example, a constraint in the UAV strike mission mentioned previously could be ``no fire munition command shall be issued when friendly forces are in the targeting area." These constraints may or may not already be reflected in the operational procedures of the system, but if not, then they can be used to guide any changes made to better ensure the safe operation of the cyber-physical system in a particular mission. 

Finally, we can identify causal scenarios using all of the previously defined information to determine how an unacceptable loss may occur. Using the UAV strike mission as an example, an unacceptable loss can occur when the operator issues a fire munition command when there are friendly forces within the target area because his or her sensors indicated otherwise. Such a scenario could feasibly be the result of a simple Denial-of-Service Attack on the operator's sensor. By creating these causal scenarios, we seek to determine the most likely or most damaging pathways for potential security breaches. Furthermore, creating the STPA-Sec model helps identify the most critical components, features, or functionality in a system with respect to mission success. This information can then be used to guide which cyber-security or cyber-resiliency measures are implemented in the future. 

\section{Application of Approach}
This methodology is demonstrated by analyzing an intelligence-gathering mission using a small reconnaissance UAV. The results and outputs of the analysis are included in this section. 

\subsection{War Rooming for the UAV mission}
For the UAV mission, the War Room activity included two commanders who are responsible for planning and organizing the mission, two system designers with technical expertise for the UAV, and an analyst leading the War Room exercise.

At the beginning of the exercise, all stakeholders agreed on a general mission and system description before entering discussions. The description of the mission is as follows: ``The tactical reconnaissance mission is to obtain visual information on the activities or resources, or lack thereof, of an enemy within a particular area to support other on-going or planned operations.'' The subsequent system description is defined as ``The tactical reconnaissance UAV carries an imaging payload to observe an area of interest and transmit the video feed to the appropriate decision-makers in support of other on-going operations in the region.'' 
The mission definition should highlight a need for a particular output and the system description should highlight how it can provide that output to the mission. In this case, the reconnaissance mission requires information about enemy activities in a particular area, and the UAV can provide that information through visual images by loitering over the area. 


Following the gathering of this general information, the commanders detailed the criteria for mission success and failure. Mission success would be completed via the collection and transmission of a clear, continuous video feed of the area of interest, regardless of what that video feed displayed. Mission failure would result from the failure to detect an actual threat in the area of interest. Furthermore, the military stakeholders also identified a small set of unacceptable outcomes. First and foremost, any loss of resources stemming from the lack of or inaccuracy of information collected during the reconnaissance mission is unacceptable. Additionally, loss of classified information or systems as a result of the loss of the UAV or any collateral damage that occurs as a result of the loss of the UAV, i.e. a crash, are unacceptable outcomes to the mission. 

Finally, the analyst queried the stakeholders about how some abnormal scenarios might be handled. These questions generally take the form of ``what if this happens?'' Clearly, this is not an exhaustive approach to identifying possible courses of action in abnormal scenarios, but the purpose here is to get a general idea of how the stakeholders may react to losses of functionality in the CPS. For example, the military commanders indicated that the reconnaissance mission would fail if no visual information could be collected by the UAV and that information would need to be gathered by other methods. These other methods could involve sending in a team of reconnaissance troops; so obviously it would be preferable to avoid putting human lives at risk. The system designers indicated that if the UAV lost GPS service, then the integrity of the mission would be compromised, but not necessarily result in mission failure as the inertial navigation system would take over. In general, for this particular mission, the loss of video functionality directly results in mission failure; meanwhile, loss of other functionality, such as GPS navigation, can result in mission failure, but is not necessarily an immediate result. This information helps us prioritize unacceptable losses and hazards when we build the STPA-Sec model.

\subsection{STPA-Sec Model for a UAV Reconnaissance Mission}
The first step in building the STPA-Sec model is to define the mission and the CPS in the context of its role in the mission. This system and mission was defined as follows: ``A reconnaissance UAV is a system to gather and disseminate information and/or data by means of imaging (or other signal detection) and loitering over an area of interest to contribute to accurate, relevant, and assured intelligence that supports a commander’s activities within and around an area or interest.'' This statement is effectively a combination of the War Room definitions of both the UAV and the mission it performs. 
The next step is to identify the unacceptable losses that could occur during the mission. In this case, this information comes directly from the War Room. The unacceptable losses are defined in order of priority in Table~\vref{tab:losses}. Given the tactical nature of this mission and the small size of the UAV, it is less vital to be concerned with the loss of the vehicle itself, but rather the loss of potentially key intelligence from the inability to survey the area of interest. 

\begin{table*}[!t]
\renewcommand{\arraystretch}{1.3}
\caption{Unacceptable losses for a UAV reconnaissance mission.}
\label{tab:losses}
\centering
\begin{tabular}{@{}ll@{}}
\toprule
Unacceptable Loss& Description\\
\midrule
L1 & Loss of resources, e.g., human, matériel, due to inaccurate, wrong, or absent information\\
L2& Loss of classified or otherwise sensitive technology, knowledge, or system(s)\\
L3 & Loss of strategically valuable matériel, personnel, or civilians due to loss of control of system(s)\\
\bottomrule
\end{tabular}
\end{table*}

\begin{table*}[!t]
\renewcommand{\arraystretch}{1.3}
\caption{Hazards that can cause unacceptable losses.}
\label{tab:hazards}
\centering
\begin{tabular}{@{}l p{4cm} p{4.5cm}@{}}
\toprule
Hazard& Worst-case Environment& Associated Losses\\
\midrule
H1---Absence of information& Imminent threat goes undetected& L1: Manpower, matériel, territory, etc.\\ 
H2---Wrong or inaccurate information& Threat is incorrectly identified or characterized& L1: Manpower, matériel, territory, etc.\\
H3---Loss of control in unacceptable area& UAV is lost in enemy territory and suffers minimal damage in crash/landing& L2, L3: Compromise of critical systems, intelligence, and/or other potentially classified information or technology\\
\bottomrule
\end{tabular}
\end{table*}

\begin{table*}[!t]
\renewcommand{\arraystretch}{1.3}
\caption{Hazard Actions.}
\label{tab:actions}
\centering
\begin{tabular}{@{}p{4.5cm} p{3cm} p{3cm} p{3cm} p{3cm}@{}}
\toprule
Control Action & Not Providing\par Causes Hazard & Providing\par Causes Hazard & Incorrect Timing\par or Order & Stopped Too Early\par or Applied Too Long \\
\midrule
CA 1.1\par Designate area of interest & H1: No information collected & H1, H2: Area is wrong or will not provide needed information  & H1, H2: Area designated is no longer of use & H1, H2: Area would be useful at another time \\
CA 1.2\par Specify surveillance target & H1, H2: Surveillance is not focused and provides too little or too much information  & H1, H2: Target is wrong or does not provide needed information & H1, H2: Target is no longer of interest or does not provide needed information & H1: Needed information occurs before or after surveillance \\
CA 1.3\par Indicate type of intelligence needed & H1, H2: Gather too much or too little data to be useful & H1, H2: Intelligence type is appropriate for what is needed & H1, H2: Type of intelligence collected at wrong time, i.e., SIGINT during time with no signals & H1: Miss desired type of intelligence \\
CA 1.4\par Create rules of flight or engagement & H3: UAV strays into inappropriate area & H1, H2: UAV cannot collect needed information  & H1, H2: Needed information not collected & H1, H2: Needed information not collected \\
\bottomrule
\end{tabular}
\end{table*}

\begin{table*}[!t]
\renewcommand{\arraystretch}{1.3}
\caption{Safety Constraints.}
\label{tab:safety}
\centering
\begin{tabular}{@{}p{4.5cm} p{8cm}@{}}
\toprule
Control Action & Safety Constraint\\
\midrule
CA 1.1\par Designate Area of Interest & The mission planner shall  clearly define the area of interest to be congruent with or a superset of the area of operations of any current or future mission for which reconnaissance is needed\\
CA 1.2\par Specify Surveillance Target & The mission planner shall indicate a specific target for the reconnaissance\\
CA 1.3\par Indicate type of intelligence needed & The mission planner shall designate a specific type of intelligence that the mission is going to collect\\
CA 1.4\par Create Rules of Flight or Engagement & The mission planner shall indicate a specific set of rules of engagement to prevent confusion at the operational level\\
\bottomrule
\end{tabular}
\end{table*}

Next, following the work flow in Figure\vref{fig:concept}, we identify a set of hazards, the worst-case environment for that hazard to occur in, and the unacceptable loss that could result from that hazard. These hazards are defined in Table\vref{tab:hazards}. The three hazards listed were determined to create the greatest threat of resulting in an unacceptable loss. The War Room indicated that the information collected during this mission is critical for mission success; therefore, the top hazards relate to the absence or unreliability of information. 

\begin{figure}[!t]
  \centering
  \input{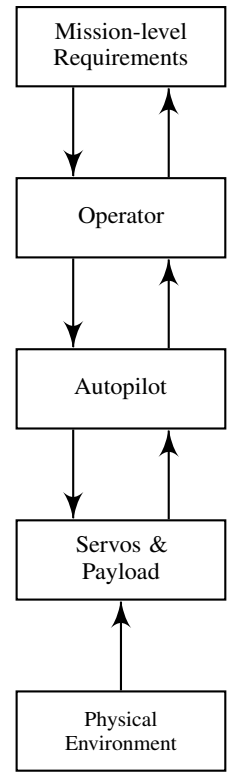}
  \caption{A hierarchical controls model that defines the expected service of a UAV. Each level is defined by a generic control structure. Inadequate control in each level can cause an adversarial action to degrade the expected service and produce a hazardous state.}
  \label{fig:hierarchy}
\end{figure}

The next step is to identify the generic control actions that can be taken at different functional levels in the system in order to provide causal paths to a particular hazard. For this mission, there were five functional levels defined: Mission-level requirements or plans, the human operator or pilot, the UAV autopilot system, the control servos and imaging payload, and the physical environment in which the UAV operates. At each level, there are a set of control actions that can be taken to influence the behavior of the surrounding levels, apart from the physical environment, which only provides disturbances to the control structure. This generic structure is represented in Figure\vref{fig:hierarchy}. For example, at the mission-level a commander will designate the area of interest for the reconnaissance mission, which feeds into the actions that the pilot takes to satisfy that requirement, and so on. For each control action, there is a scenario in which one of the four types of unsafe control actions creates at least one of the hazards identified in Table\vref{tab:hazards}. A subset of these control actions and circumstances are outlined in Table\vref{tab:actions}. 

Now that we have identified the control actions available in the system and the conditions under which they create hazardous scenarios, we can identify a set of constraints that can be applied to the behavior of the system to limit the possibility of a hazardous scenario leading to an unacceptable loss. The constraints defined for the control actions outlined in Table\vref{tab:actions} is presented in Table\vref{tab:safety}. 

In addition to the constraints that should be applied on the system, analysis of the STPA-Sec model identifies areas that should receive the most attention in order to increase security and resiliency against cyber attacks that can produce unacceptable mission outcomes. For the UAV reconnaissance mission identified in this example, the most pressing unacceptable outcome relates to military commanders not receiving vital information about potential enemy activity within an area of interest. In this case, the integrity of the video feed coming from the UAV should receive top priority. Developing and evaluating measures for ensuring integrity of the video feed (or assuring that the system can identify when integrity has been lost) is outside of the scope of this paper.

\section{Related Works}
Cybersecurity generally follows a software-oriented perspective. Mead and Woody present state-of-the-art methods for software assurance in~\cite{mead_security}, and focus on integrating security earlier in the acquisition and development cycle of software. Furthermore, Mead, Morales, and Alice describe an approach that uses existing malware to inform the development of security requirements in the early stages of the software lifecycle~\cite{mead_malware}. This approach seeks a similar product to the one presented in this paper; however, it follows the standard, bottom-up approach of identifying threats and generating solutions based on those threats. These techniques work well for IT software systems, yet are insufficient for cyber-physical systems. Hu asserts in~\cite{hu_book} that these cybersecurity approaches are not effective for CPS as an attack on a physical system is not necessarily detectable or counteracted by cyber systems. Burmester et al define a threat modeling framework specifically for CPS that takes into account the physical component of CPS that many other methods do not~\cite{model_analysis}, yet this approach still relies on historical threats to identify vulnerabilities. 

STPA-Sec~\cite{young_systems_2013}, however, aims to reverse the tactics-based bottom up approach of other cybersecurity methodologies. While we seek to address the same issue, our implementation of STPA-Sec differs in two key areas. First, the implementation presented by Young and Leveson is a methodology to be performed by analysts on their own, whereas our implementation is informed explicitly by mission and system stakeholders via the concept of the War Room. This aids the STPA-Sec analysis by minimizing the chance of outputs not matching the perspectives and experiences of the stakeholders. Second, the approach presented in this paper introduces a mission-aware viewpoint to the STPA-Sec analysis. That is, one could have the exact same CPS in a completely different mission context, and would potentially want to choose different security solutions. The incorporation of the mission into the analysis scopes the security problem above the cyber-physical system level, which both opens up possibilities for potential vulnerability solutions, and motivates the choice of security or resiliency-based solutions.

\section{Discussion \& Conclusions}
This paper presented a systems approach to augmenting security and resiliency in cyber-physical systems. This framework is based on a top-to-bottom identification of unacceptable losses or outcomes to a particular mission that the CPS performs and examines how the paths to those outcomes can be avoided. We have shown an application of this approach to a hypothetical tactical reconnaissance mission using a small UAV and generated a set of constraints that should be present in the behavior of this example system to avoid pathways to unacceptable outcomes. In addition, this approach identifies the areas most critical to mission success as starting points for future implementations of security or resiliency solutions. 

\balance

A future direction based on the findings of this work includes implementing the identified system constraints on model and formally checking that they can avoid unacceptable losses to the mission. Additionally, this work could be extended by closing the loop and testing security or resiliency solutions' effects on the behavior of the system in its mission. This would allow security and resiliency solutions to be evaluated based on their cost, complexity of implementation, and effectiveness at preventing unacceptable mission outcomes. 

Through this work, we have identified an approach to reversing the traditional bottom-up nature of other security methodologies based on a mission-aware viewpoint. This approach recognizes that security is a hard and complex problem, but seeks to manage the costs and complexity of increasing security and resiliency by focusing on avoiding unacceptable losses rather than reacting to threats as they appear. 

\section*{Acknowledgments}

This research is based upon work supported by the Systems Engineering Research Center under Award No. 2016-TR-156.

\bibliographystyle{IEEEtran}
\bibliography{manuscript}

\begin{thebibliography}{10}
\providecommand{\url}[1]{#1}
\csname url@samestyle\endcsname
\providecommand{\newblock}{\relax}
\providecommand{\bibinfo}[2]{#2}
\providecommand{\BIBentrySTDinterwordspacing}{\spaceskip=0pt\relax}
\providecommand{\BIBentryALTinterwordstretchfactor}{4}
\providecommand{\BIBentryALTinterwordspacing}{\spaceskip=\fontdimen2\font plus
\BIBentryALTinterwordstretchfactor\fontdimen3\font minus
  \fontdimen4\font\relax}
\providecommand{\BIBforeignlanguage}[2]{{%
\expandafter\ifx\csname l@#1\endcsname\relax
\typeout{** WARNING: IEEEtran.bst: No hyphenation pattern has been}%
\typeout{** loaded for the language `#1'. Using the pattern for}%
\typeout{** the default language instead.}%
\else
\language=\csname l@#1\endcsname
\fi
#2}}
\providecommand{\BIBdecl}{\relax}
\BIBdecl

\bibitem{Equifax}
B.~Krebs, ``Equifax hackers stole 200k credit card accounts in one fell
  swoop,'' \url{https://perma.cc/64EM-SAZF}, accessed: 2017-16-09.

\bibitem{Havex}
``Alert ({ICS-ALERT-14-176-02A}) ics focused malware (update a),''
  \url{https://ics-cert.us-cert.gov/alerts/ICS-ALERT-14-176-02A}, accessed:
  2016-12-05.

\bibitem{Stuxnet}
``Advisory ({ICSA-10-201-01C}) usb malware targeting siemens control software
  (update c),'' \url{https://ics-cert.us-cert.gov/advisories/ICSA-10-201-01C},
  accessed: 2017-10-05.

\bibitem{leveson_engineering_2012}
N.~Leveson, \emph{\BIBforeignlanguage{English}{Engineering a safer world:
  systems thinking applied to safety}}.\hskip 1em plus 0.5em minus 0.4em\relax
  Cambridge, Mass.: The MIT Press, 2012.

\bibitem{frola_system_1984}
F.~Frola and C.~Miller, ``System safety in aircraft management,''
  \emph{Logistics Management Institute, Washington DC}, 1984.

\bibitem{strafaci_what_2008}
A.~Strafaci, ``What does bim mean for civil engineers,'' \emph{CE News,
  Tranportation}, 2008.

\bibitem{young_systems_2013}
W.~Young and N.~Leveson, ``Systems thinking for safety and security,'' in
  \emph{Proceedings of the 29th Annual Computer Security Applications
  Conference}, ser. {ACSAC} '13.\hskip 1em plus 0.5em minus 0.4em\relax {ACM},
  pp. 1--8.

\bibitem{fleming_systems_2017}
C.~H. Fleming, ``Systems theory and a drive towards model-based safety
  analysis,'' in \emph{2017 {Annual} {IEEE} {International} {Systems}
  {Conference} ({SysCon})}, Apr. 2017, pp. 1--5.

\bibitem{mead_security}
N.~R. Mead and C.~C. Woody, \emph{Cyber Security Engineering: A Practical
  Approach for Systems and Software Assurance}.\hskip 1em plus 0.5em minus
  0.4em\relax Addison-Wesley, 2017.

\bibitem{mead_malware}
N.~R. Mead, J.~A. Morales, and G.~R. Alice, ``A method and case study for using
  malware analysis to improve security requirements,'' \emph{International
  Journal of Secure Software Engineering}, vol.~6, pp. 1--23, January 2015.

\bibitem{hu_book}
F.~Hu, \emph{Cyber-Physical Systems: Integrated Computing and Engineering
  Design}.\hskip 1em plus 0.5em minus 0.4em\relax Boca Raton, FL: CRC Press,
  Inc, 2013.

\bibitem{model_analysis}
M.~Burmester, E.~Magkos, and V.~Chrissikopoulos, ``Modeling security in
  cyber-physical systems,'' \emph{International Journal of Critical
  Infrastructure Protection}, vol.~5, pp. 118--126, December 2012.

\end{thebibliography}
\end{document}